\documentclass[peerreview]{IEEEtran}
\IEEEoverridecommandlockouts
\usepackage{cite}
\usepackage{amsmath,amssymb,amsfonts}
\usepackage{algorithmic}
\usepackage{graphicx}
\usepackage{textcomp}
\usepackage{xcolor}
\def\BibTeX{{\rm B\kern-.05em{\sc i\kern-.025em b}\kern-.08em
    T\kern-.1667em\lower.7ex\hbox{E}\kern-.125emX}}
\begin{document}

\title{Using Data Mining for Infrastructure and Safety Violations Discovery in Cities\\
}

\author{\IEEEauthorblockN{Doron Laadan, Eyal Arviv, and Michael Fire}
\IEEEauthorblockA{\textit{Department  of Software and Information Systems Engineering} 
\textit{Ben-Gurion University of the Negev}
Beer-Sheva, Israel \\
\{laadan,eyalar,mickyfi\}@post.bgu.ac.il}
}
\maketitle

\begin{abstract}
    In city planning and maintenance, the ability to quickly identify infrastructure violations - such as missing or misplaced fire hydrants - can be crucial for maintaining a safe city; it can even save lives. In this work, we aim to provide an analysis of such violations, and to demonstrate the potential of data-driven approaches for quickly locating and addressing them. We conduct an analytical study based upon data from the city of Beer-Sheva’s public records of fire hydrants, bomb shelters, and other public facilities. The results of our analysis are presented along with an interactive exploration tool, which allows for easy exploration and identification of the different facilities around the city that violate regulations.
\newline \textbf{key-words:} Data-mining, Data-science, Smart Cities


\end{abstract}

\section{Introduction}
    As it has with so many other fields, the data science revolution is changing the field of city planning; in recent years, experts in the field have begun to fully utilize the data-rich environment that city planning provides\cite{miller2009geographic}. This data-rich environment is fueled by both the public and private sectors, which create, process, and store digital geographic datasets for everything from cities to entire countries\cite{souza2019data}. This explosion of different data sources has enabled applied data mining techniques to uncover useful and previously undetected patterns that may improve the population’s safety and quality of life. One example of such work is CityGuard \cite{ma2017cityguard}, which detects safety conflicts in smart cities.


In city planning, figuring out the ideal location for placing a fire hydrant or finding the locations of different infrastructure safety violations is crucial; historically, this work has mostly been done by city planners and government offices \cite{corburn2009toward}. Using data mining, it is possible to quickly identify suitable areas for various facilities and find locations that violate safety regulations, such as a health center or daycare center with no fire hydrant nearby.


In this study, we aim to utilize this geo-data to identify crucial infrastructure violations; we present a method which utilizes a city’s open-source datasets to do so. We evaluated our method using data from the city of Beer-Sheva. In this particular case, we utilized our method to identify missing fire hydrants and bomb shelter placement in the different neighborhoods in Beer-Sheva.


Our method utilizes data of the locations of various facilities - such as health centers, religious centers, community centers, and more - as well as the geo-data of fire hydrants and bomb shelters in the different neighborhoods of the city into a graph (see Figure \ref{fig:FH10}). Then, we analyze the graph and detect areas and facilities with a large number of missing infrastructures. By using graph centrality algorithms, we can identify key central points in the city for the placement of fire hydrants or bomb shelters (i.e., locations where adding or removing them can have a significant effect).


Using our analysis, we can point out disparities between different areas of the city regarding infrastructure and safety violations. For example, our analysis uncovered a significant difference between the different neighborhoods of Beer-Sheva regarding compliance with fire safety regulations. Also, our analysis can assist decision-makers to identify regions, which lack crucial infrastructures. Lastly, we provide a simple interactive tool to explore this analysis. 

The remainder of this paper is organized as follows: in Section II, we provide a brief overview of related work; in Section III, we describe the methods that were used in our study; in Section IV, we present the results that we obtained; and finally, in Section V, we give our conclusions from this study, and we offer future research directions.
 
\section{Related Work}
    \subsection{Geographical Data Mining}
In recent years due to the continuing work from government agencies, private sectors, and scientific projects to collect and store geographical data, the field of geographic research was able to move into a data-rich environment \cite{mennis2009spatial}. The move to this data-rich environment enabled researches to use data mining tools to explore and find unique patterns in geographical data.

Geographic data mining and knowledge discovery (GKD) is a special case of Knowledge Discovery in Databases (KDD) \cite{miller2009geographic}. 
The unique property of GKD is that up to four dimensions in the data can create a measurement framework for all other dimensions, the most common being the topology and geometry consistent with the Euclidean distance. This framework can affect the attributes and patterns that can be analyzed in the data \cite{miller2003representation}.

\subsection{Smart Cities}
Smart cities (SC) are designed to improve citizen welfare, boost economic development, and help people build sustainable services \cite{souza2019data}. The defining  characteristics of a smart city are sustainability, resilience, governance, enhanced quality of life, and intelligent management of resources and facilities. To facilitate these characteristics, smart cities require the employment of a wide range of  computational methods such as data mining (DM) and machine learning (ML)\cite{yin2015literature}.
Both DM and ML can use the vast amount of data a smart city holds in its different applications to enhance smart city services further \cite{al2015applications}.
Such services can use data mining to improve road safety \cite{fire2012data} or assist the city’s crowd-wise policy and decision making\cite{giatsoglou2016citypulse}. We will use GKD to help identify infrastructure and safety violations in cities.  

\section{Methods}
    
In this section, we will articulate the details of our analytical approach. Given a city, we project facilities, locations, and objects of interest (such as fire hydrants or bomb shelters) into a weighted graph. We define a weighted graph $G := <V, E>$ as a set of vertices $V$, and a set of links between vertices $E$. We construct $G$ in such a way that each vertex $v \in V$ corresponds to different facilities or objects in the city. For instance, a specific community center may be represented by a vertex $v \in V$. Then, we define a set of links $E := \{(u,v) | {u,v} \in V,  \text{w(u,v)} \leq th\}$, where each weighted link $e = (u,v)$ represent a link between facilities and objects. We define each link's weight, $w(u,v)$, to be equal to the geographic aerial distance between two vertices. We set $th$ to be a predefined threshold parameter. For example, for any given city, we can generate a graph $G$ to be a graph that consists of all daycare centers and fire hydrants in the city where links are present only between those daycare centers and fire hydrants, which are less than 100 meters from each other.

By analyzing $G$, we can obtain different insights about the city in question. For example, if we construct $G$ to be a graph that connects daycares and fire hydrants we can identify facilities without fire hydrants nearby, which violate fire department safety regulations, by retrieving isolated vertices in $G$. Additionally, by using graph centrality measures such as degree centrality\cite{getoor2005link}, we can locate central objects in the city, such as central fire hydrants that affect a large number of facilities. Using this knowledge, city official can prioritize these central objects for maintenance work. Our analysis can also suggest places suitable for a new bomb shelter or fire hydrant.



In order to construct the dataset for our analysis, we collected data from the “Smart7 OpenData” portal,\footnote{https://www.beer-sheva.muni.il/OpenData/Pages/default.aspx}  which provided access to Beer-Sheva city data. The geo-data includes the geographical coordinates (longitude and latitude) of different facilities in Beer-Sheva, such as community centers, daycare centers, as well as the locations of bomb shelters and fire hydrants.


Next, we defined the boundaries of each neighborhood in the city so that our analysis could take them into account. Beer-Sheva has 15 neighborhoods.\footnote{The  neighborhoods names are Alef, Bet, Gimel, Dalet, Hei, Vav, Tet, Ramot, Down-Town, Yod-Alef, Old-Town, Ashan, Noi-Beka, Darom, and Nahot.} Since this information is not recorded in the Smart7 OpenData portal, we integrated the boundaries of these neighborhoods into our model using data from the city of Beer-Sheva’s website and Google Maps. Having done so, we constructed a graph $G_H$, such that each facility and fire hydrant in the data was stored as a vertex in the graph with its geo-location (altitude and longitude). Next, we calculated the geographic aerial distance between each facility vertex and each fire hydrant vertex and used the result as the weight of the link between the two vertices. Lastly, we filtered the edges in the graph so that only those which fell below the previously defined threshold parameter $th$ remained. We repeated the process and constructed an additional graph $G_S$, which used the locations of bomb shelters rather than those of fire hydrants. locations of bomb shelters.

\begin{figure}[h]
	\centering
	\includegraphics[scale=0.65]{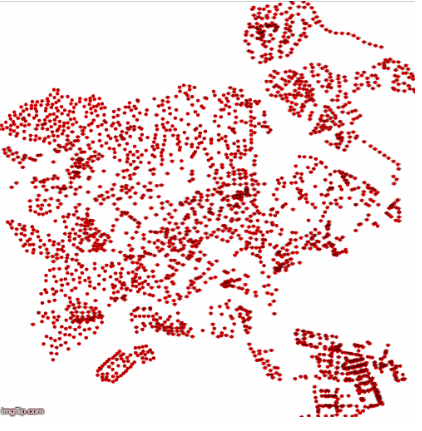}
	\caption[]{A graph where each node is a fire hydrant and there exists an edge between nodes if the distance between them is less than 10 meters}
	\label{fig:FH10} 
\end{figure}

\begin{figure*}[t!]
	\centering
	\includegraphics[scale=0.65]{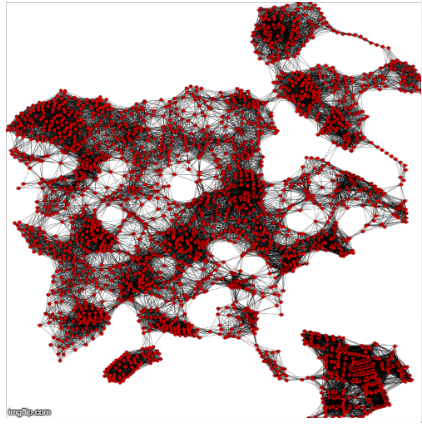}
	\caption[]{A graph where each node is a fire hydrant and there exists an edge between nodes if the distance between them is less than 500 meters}
	\label{fig:FH500} 
\end{figure*}

An example of a simple graph that was built using our method using only the fire hydrant objects (essentially, the fire hydrant deployment in Beer-Sheva) can be seen in Figs \ref{fig:FH10} and \ref{fig:FH500}, which show the fire hydrant graphs when the threshold parameter is set at 10 meters and 500 meters, respectively. It is interesting to see that the topology of the city, as seen in Fig. 3, is reflected in the graph itself.

\section{Result Analysis}
    Our analysis includes the next facilities: community-centers, daycare-centers, gas stations, educational institutions, health clinics, sport-centers, and synagogues. Overall there were 1000 data points representing facilities across the city used in this work. We had 2596 and 265 data points representing fire hydrants and bomb shelters, respectively, across the city.

\subsection{Fire Hydrants Analysis}
According to the Beer-Sheva city fire department regulations, each facility must be in no more than 30 meters away from the closest fire hydrant. Using this as our threshold, we built our graph $G_H$ using the previously articulated method and identified the isolated nodes.

\begin{figure}[h!]
	\centering
	\includegraphics[width=0.4\textwidth]{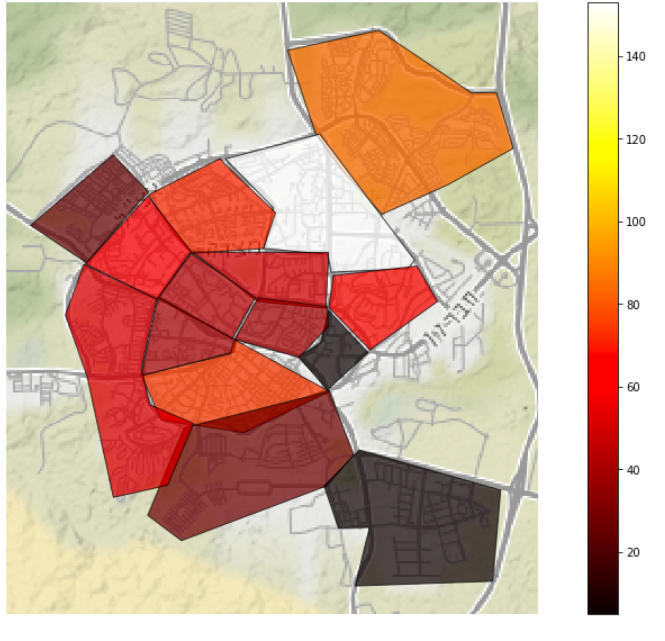}
	\caption[]{A heatmap of Beer-Sheva neighborhoods according to the amount of violations.}
	\label{fig:city_heat} 
\end{figure}

\begin{figure}[ht!]
	\centering
	\includegraphics[width=0.55\textwidth]{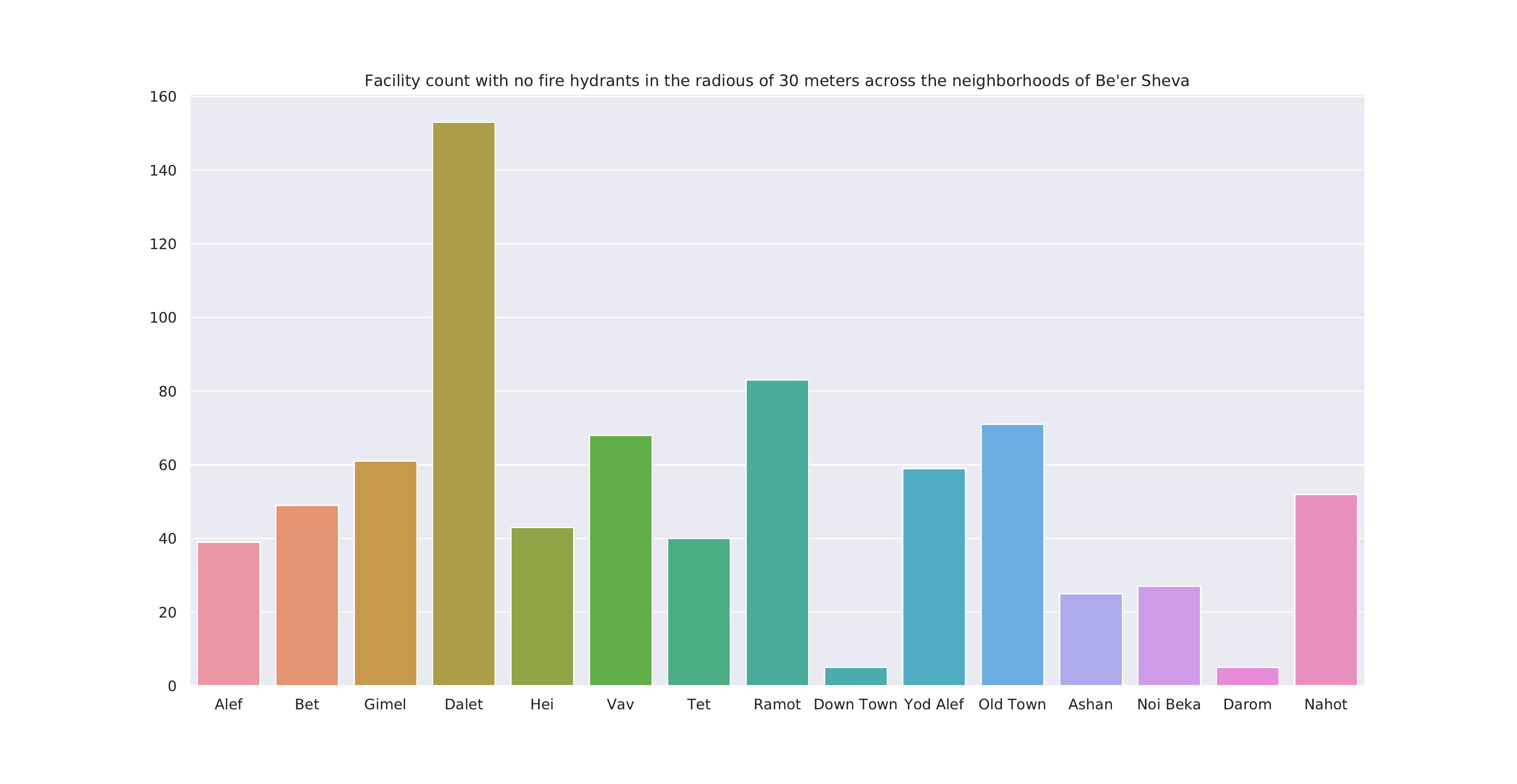}
	\caption[]{A bar plot of Beer-Sheva neighborhoods according to the amount of violations.}
	\label{fig:FH_bar} 
\end{figure}

Using our analysis, we identified over 800 facilities that break fire department regulations in the city and, using heatmaps, we identified neighborhoods that have the most violations in them. In Fig. \ref{fig:city_heat} and Fig.~\ref{fig:FH_bar} we can see a heat map and a bar plot of the city’s various neighborhoods, both of which are organized according to the number of fire regulation violations found in each neighborhood.

It is interesting to note that no neighborhood is free of violations. That being said, it would appear that the Dalet neighborhood is the neighborhood with the most violations, while the fewest number of infractions are found in the Down-Town and Darom neighborhoods.  The Down-Town  neighborhood is considered a central hub of the city, and therefore a low amount of violations are to be expected. The Darom area is mainly an industrial district, so the low number of violations can be explained by the relative lack of facilities. Interestingly, the Ramot neighborhood, which is a relatively new neighborhood in the city, has the second most violations. It is important to note that our data was last updated in September 2019 and, as is the case with all open source data, may have missing information or some errors.

\begin{figure}[ht!]
	\centering
	\includegraphics[width=0.6\textwidth]{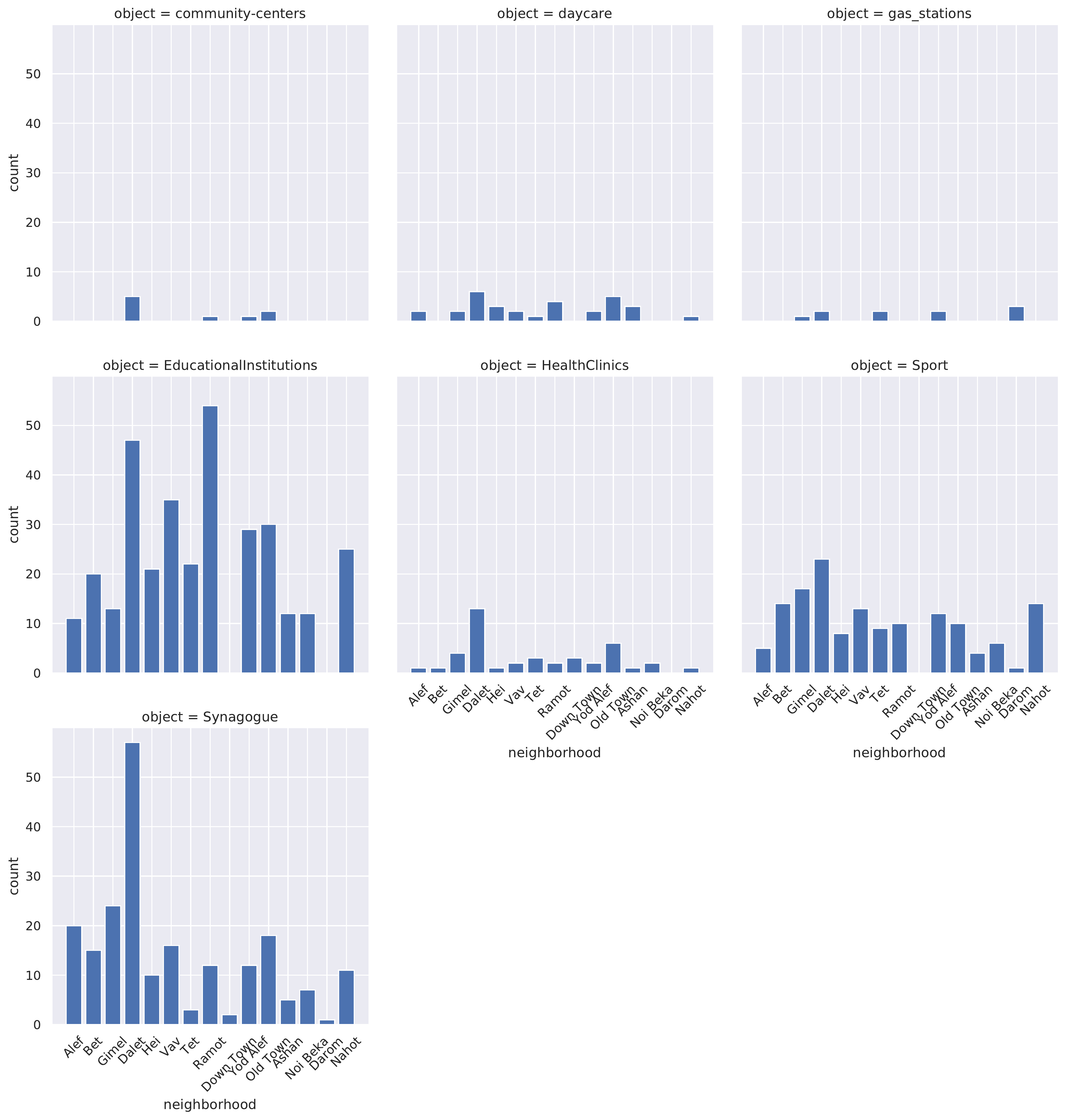}
	\caption[]{A bar plot of Beer-Sheva facilitates according to the amount of violations in different neighborhoods.}
	\label{fig:bar_faciletis} 
\end{figure}

Next, we will explore the violations across different facilities. Fig.~\ref{fig:bar_faciletis} shows the number of breaches according to the type of facility in each neighborhood. Upon close analysis, one may observe that, amongst the different kinds of facilities explored in this work, religious and educational centers have the most violations. This could be seen as something of a red flag since these locations tend to have a high amount of people near them most hours of the day. There is the possibility that these facilities have some sort of a built in fire prevention systems and as such does not require fire hydrants nearby.

\begin{figure}[ht!]
	\centering
	\includegraphics[scale=0.48]{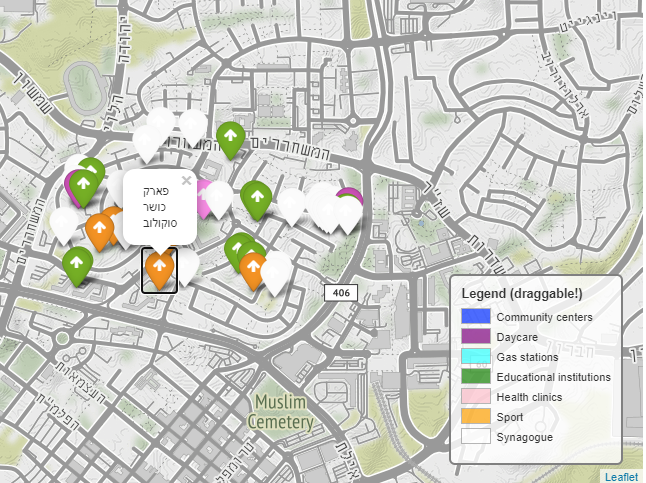}
	\caption[]{A snippet from our tool showing the names and locations of facilities in 'Alef' which violate regulations (threshold is set to 30 meters as the law).}
	\label{fig:snippet} 
\end{figure}

Finally, we provide a snippet from our interactive tool , which allows us to quickly explore the data and results by selecting the specific neighborhood(s) and facilities from the dataset that we want to study. Fig.~\ref{fig:snippet} demonstrates the results of selecting all of the facilities in the Alef neighborhood at a threshold of 30 meters (in accordance with fire department regulations). The interactive tool also provides the names and locations of facilities that violate regulations.

\subsection{Bomb Shelters Analysis}
We conducted a similar analysis of the city regarding bomb shelter regulations. According to the Beer-Sheva city fire department, each facility must be within no more than 50 meters distance from the nearest bomb shelter. Using this as our threshold, we built our graph $G_S$ using the previously described method and identified the isolated nodes.

\begin{figure}[h!]
	\centering
	\includegraphics[scale=0.55]{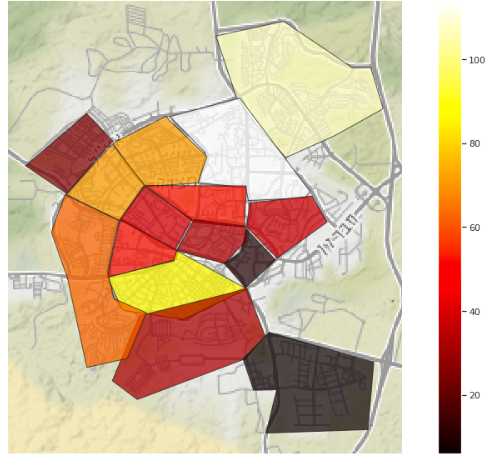}
	\caption[]{A heatmap of Beer-sheva neighborhoods according to the amount of violations}
	\label{fig:city_heat_sh} 
\end{figure}

\begin{figure}[h!]
	\centering
	\includegraphics[width=0.75\textwidth]{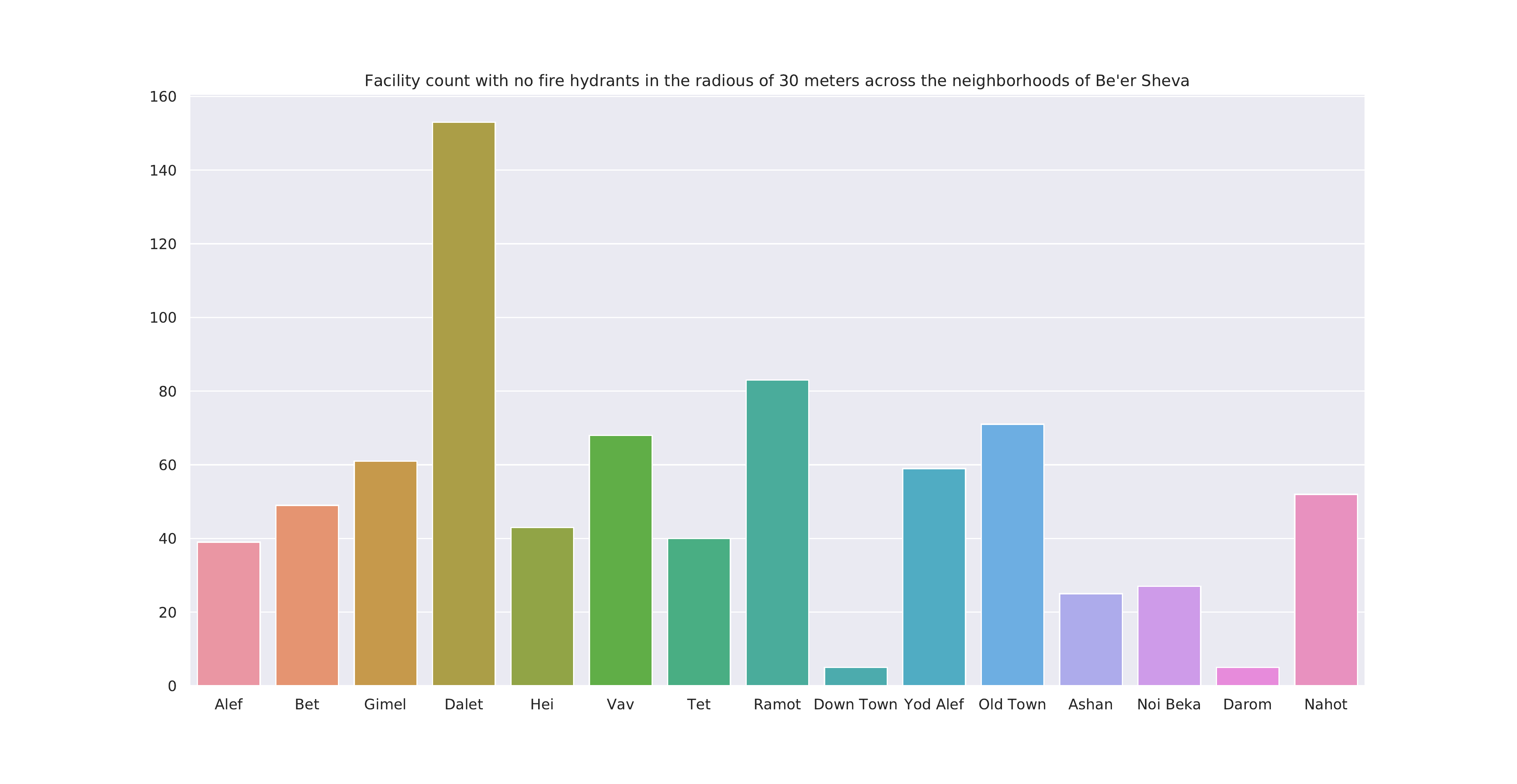}
	\caption[]{A bar plot of Beer-sheva neighborhoods according to the amount of violations}
	\label{fig:sh_bar} 
\end{figure}

Fig. \ref{fig:city_heat_sh} and Fig. \ref{fig:sh_bar} show a heatmap and a bar plot of the city according to the different neighborhoods and the number of facilities violating safety regulations that each area has.

The results that we obtained reveal a higher frequency of violations regarding bomb shelters than we discovered in our analysis of fire hydrants. The relative distribution of violations amongst neighborhoods, however, appears largely the same; it would appear that the Dalet and Rammot neighborhoods have the highest number of violations, and that the Down-Town and Darom neighborhoods have the fewest.

\begin{figure}[h!]
	\centering
	\includegraphics[width=0.7\textwidth]{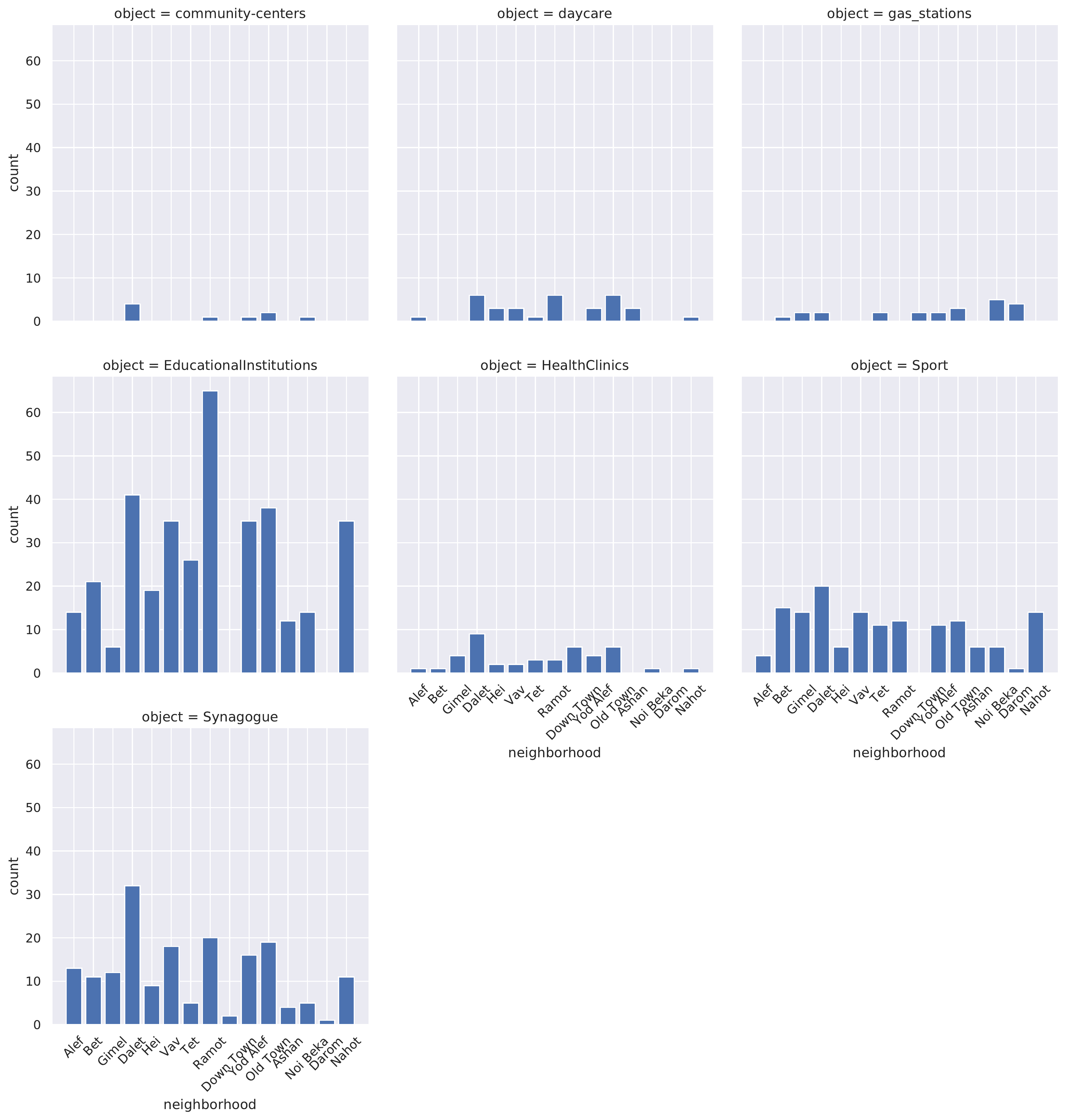}
	\caption[]{A bar plot of Beer-Sheva facilitates according to the amount of violations in different neighborhoods}
	\label{fig:bar_faciletis_sh} 
\end{figure}

Fig.~\ref{fig:bar_faciletis_sh} shows the number of breaches according to the type of facility in each neighborhood. From figure, we can again observe that religious and educational centers have the most violations. In the particular case of educational centers, this high frequency of violations may be a reflection of the legal requirement that each center must contain a bomb shelter within the facility, eliminating the need to rely on public shelters. 

\section{Conclusions and Future Directions}
    In this work, we used data mining to analyze the geographical data of Beer-Sheva to quickly identify areas and facilities that violate fire and safety regulations by projecting the the geo-data into a graph. We were able to find hundreds of violations in different parts of the city across several types of facilities. Our analysis shows the disparities between different parts of the city; additional meaningful disparities could be explored using our method by making use of other available data, such as economic status and the like.

For future works, we plan on creating an automated system that can highlight violations and recommend suitable areas to build new facilities or ideal areas to prioritize maintenance work to city officials.

\section*{ACKNOWLEDGMENT}
We would like to thank Michael Hiestand for editing and proofreading this article.
\bibliographystyle{IEEEtran}
\bibliography{bibtex.bib}

\vspace{12pt}

\end{document}